\documentclass[%
reprint,
onecolumn,
amsmath,amssymb,
aps,
floatfix,
notitlepage,
]{revtex4-2}

\usepackage{amssymb,amsmath}
\usepackage{amsmath}
\usepackage{graphicx}
\usepackage{subcaption}
\usepackage{gensymb}
\graphicspath{{fig/}}
\usepackage[colorinlistoftodos]{todonotes}
\usepackage{xcolor}
\usepackage{xfrac}
\usepackage[normalem]{ulem}
\usepackage{tikz}
\usepackage{siunitx}
\usepackage{float}
\usepackage{empheq}

\usepackage{hyperref}
\hypersetup{
	colorlinks,
	citecolor=blue,
	linkcolor=black,
	urlcolor=black
}

\definecolor{darkolivegreen}{rgb}{0.33, 0.42, 0.18}

\usepackage{setspace}
\usepackage[margin=2.5cm]{geometry}

\definecolor{corange}{HTML}{ff7d0c}
\definecolor{cblue}{HTML}{1e75b3}

\begin{document}

\title{Second-order stochastic modeling of particle resuspension:  macroscopic degeneracy and anomalous pre-detachment transport}

\author{David Ben-Shlomo}
\affiliation{Department of Chemical Engineering, Ben-Gurion University of the Negev, Beer-Sheva 8410501, Israel}
\affiliation{Department of Applied Mathematics, Israel Institute for Biological Research, Ness-Ziona 7410001, Israel}

\author{Ronen Berkovich}
\email[Corresponding author: Ronen Berkovich, ]{berkovir@bgu.ac.il}
\affiliation{Department of Chemical Engineering, Ben-Gurion University of the Negev, Beer-Sheva 8410501, Israel}

\author{Eyal Fattal}
\email[Corresponding author: Eyal Fattal, ]{eyalfattal@yahoo.com}
\affiliation{Department of Applied Mathematics, Israel Institute for Biological Research, Ness-Ziona 7410001, Israel}



\begin{abstract}
Particle resuspension models are commonly evaluated using the macroscopic resuspended fraction, although this integrated observable may conceal the temporal dynamics leading to detachment. Here, a second-order Markovian Lagrangian stochastic model is developed by augmenting the angular-velocity state with a finite-correlated tangential acceleration. The model is examined over turbulent channel flows with ($Re_\tau \in [60, 430]$) and compared with an established first-order formulation. The two models produce nearly overlapping resuspended fractions, revealing a macroscopic degeneracy between distinct stochastic descriptions. Multiscale trajectory statistics break this degeneracy. The acceleration-augmented formulation changes the short-time regularity from $S_2(\tau) \propto \tau$ to $S_2(\tau) \propto \tau^2$, sustains angular-velocity correlation, and produces stronger directional asymmetry and heavier increment tails. The survivor-conditioned mean square displacement exposes a Reynolds-number-dependent anomalous-transport window, in which the attached-particle ensemble grows more rapidly than the diffusive reference and locally approaches ballistic and super-ballistic scaling before crossing toward diffusion-like transport. This finite-time pathway records how particles approach detachment and is compressed out of the macroscopic resuspended fraction. At $Re_\tau \approx 60$, the trajectories enter a distinct low-Reynolds-number statistical state characterized by converging velocity and acceleration decorrelation and near-Gaussian increments. These results establish multiscale trajectory statistics as essential discriminants between stochastic resuspension models and identify finite acceleration correlation as a source of dynamical information beyond macroscopic detachment kinetics.
\end{abstract}
\maketitle


\section{Introduction}\label{sec:intro}

The detachment and resuspension of micro-particles from surfaces by turbulent flows is an important transport phenomenon across a wide spectrum of scientific and industrial domains, ranging from environmental pollutant dispersion and aerosol epidemiology to nuclear reactor safety and semiconductor manufacturing \cite{henryParticleResuspensionChallenges2023,ziskindParticleResuspensionSurfaces2006}. Despite decades of intensive research, accurately predicting particle resuspension rates remains a formidable theoretical and computational challenge. This difficulty arises primarily from the complex, multi-scale nature of near-wall turbulence and its non-linear coupling with surface adhesion forces \cite{henryParticleResuspensionChallenges2023}.

In wall-bounded turbulent flows, particles residing within the viscous sublayer ($y^{+}<5$, where $y^{+} = yu_{\tau}/\nu$ denotes the dimensionless wall-normal coordinate, \(u_{\tau}\) is the friction velocity and \(\nu\) is the kinematic viscosity) and the buffer layer ($5<y^+<30$) are subjected to intense, aerodynamic drag and lift forces driven by coherent near-wall turbulent structures, such as streamwise velocity streaks, sweeps, and ejections. These structures were shown to penetrate the viscous sublayer, practically reaching the wall surface \cite{shengBufferLayerStructures2009,vanhoutSpatiallyTemporallyResolved2013,soltaniDirectNumericalSimulation1995}. Particularly, for particles fully immersed within the viscous sublayer, particle movement is dominated by rolling motion along the wall under the influence of aerodynamic forcing before encountering a localized aerodynamic impulse sufficient to rupture the adhesive interaction \cite{ibrahimMicroparticleDetachmentSurfaces2004,jiangCharacterizingEffectSubstrate2008,soltaniParticleAdhesionRemoval1994}. Because near-wall turbulence is random and surface adhesion forces, governed by surface roughness characteristics, exhibit broad statistical distributions, particle resuspension can be viewed as a stochastic process. Consequently, classical deterministic threshold models are not able to capture the temporal evolution and the turbulence-driven dynamics of the resuspension process.

To account for the probabilistic nature of near-wall forcing, Lagrangian-stochastic models (LSMs) based on Langevin equations have been widely adopted to simulate particle resuspension \cite{guingoNewModelSimulation2008,henryNumericalStudyAdhesion2012,henryStochasticApproachSimulation2014,huModelingResuspensionSmall2023,fuParticleResuspensionWallbounded2013,benshlomoIntroducingSurfaceRoughness2024,ben-shlomoMarkovianAssumptionNearwall2026}. LSMs offer several distinct advantages. First, they can reconstruct empirical statistical distributions through probability density functions (PDFs). In the context of resuspension, PDFs can be utilized to characterize different random variables of the process, such as the turbulence-induced drag force. This provides key insights into the underlying physics without incurring the computational cost associated with resolving the full Navier-Stokes equations, as required in direct numerical simulations (DNS). Additionally, LSMs simplify the description of turbulence in the inertial subrange by modelling it as a memoryless Markovian process, consistent with the Kolmogorov-Obukhov (K41) theory \cite{kolmogorovLocalStructureTurbulence1941,obukhovDescriptionTurbulenceTerms1959}. This modelling paradigm has proven to be a highly effective tool across numerous cases, including highly inhomogeneous turbulent flows \cite{fattalHeterogenousCanopyLagrangianStochastic2023,shnappTurbulenceobstacleInteractionsLagrangian2020,thomsonCriteriaSelectionStochastic1987,wilsonReviewLagrangianStochastic1996}.

Although the occurrence and termination of individual high- and low-drag intervals may be approximated by Poissonian statistics \cite{ben-shlomoMarkovianAssumptionNearwall2026}, the dynamics within each interval retain temporal organization. This distinction creates an identification problem for resuspension models: a first-order formulation may reproduce the integrated resuspended fraction through calibration while encoding a different temporal route to detachment. Macroscopic agreement may therefore coexist with trajectory-level dynamical non-equivalence.

The present study exploits this macroscopic degeneracy as a controlled test of model order. Augmenting the stochastic state with acceleration introduces a finite forcing-correlation time and changes the temporal regularity of the angular velocity. The first-order formulation generates rough, nondifferentiable angular-velocity paths, whereas the acceleration-augmented formulation generates differentiable short-time trajectories \cite{innocentiLagrangianStochasticModelling2020,sawfordReynoldsNumberEffects1991,popeStochasticLagrangianModel2002}. This structural distinction is expected to propagate into transport scaling, temporal correlation, increment asymmetry, and tail statistics.

The objective of this study is to determine how this change in stochastic regularity reshapes pre-detachment particle dynamics while leaving the macroscopic resuspended fraction nearly unchanged. . The two formulations are compared across friction Reynolds numbers, $Re_\tau \in [60, 430]$, using the resuspended fraction, survivor-conditioned mean-squared displacement, velocity and acceleration autocorrelation functions, and second-, third-, and fourth-order increment statistics. The analysis tests the central hypothesis that higher-order and multiscale trajectory statistics break the degeneracy of macroscopic calibration and reveal dynamical information hidden by the resuspension fraction, RF.

Within this framework, anomalous transport refers to the finite-time, non-Fickian evolution of the survivor-conditioned mean-squared displacement. It is identified through intervals in which the mean square displacement (MSD) grows more rapidly than the linear diffusive reference and spans the ballistic and super-ballistic scaling classes shown in figure \ref{fig:MSD} \cite{zotero-item-1149,metzlerRandomWalksGuide2000}. Because this observable follows the particles that remain attached, it resolves the evolving surface-transport pathway leading toward detachment rather than only the final detached fraction.


\section{Methods}\label{sec:methods}

\subsection{Particle resuspension models}\label{sec:resus_models}

The first-order Markovian LSM, previously introduced and validated against experimental data in \cite{benshlomoIntroducingSurfaceRoughness2024}, forms the baseline for the current second-order formulation. The framework assumes particles roll in a single direction along the surface \cite{soltaniParticleAdhesionRemoval1994,ibrahimMicroparticleDetachmentSurfaces2004,ziskindParticleResuspensionSurfaces2006,henryParticleResuspensionChallenges2023}, with the trajectory governed by a moment balance, $M$, around a downstream asperity pivot point:

\begin{equation}
    M \equiv I\frac{\mathrm{d}\omega}{\mathrm{dt}} = bF_{D} + \frac{a}{2}F_{L} - r_{a}F_{a} - \frac{a}{2}mg
    \label{eq:moment_balance}
\end{equation}
where $I$ is the particle moment of inertia, approximated for a sphere using the parallel axis theorem as $I = (7/20) m d_{p}^{2}$. The particle diameter and mass are denoted by $d_{p}$ and $m$, respectively; $g$ is the gravitational acceleration; $\omega$ is the angular velocity; $a$ is the distance between consecutive surface asperities; $r_a$ is the lever arm of the adhesion force; and $b = \sqrt{(d_{p}/2)^{2} - (a/2)^{2}}$.
The aerodynamic drag force is evaluated using Stokes' formulation, $F_D = 6\pi\rho_f\nu(d_p/2)\langle u \rangle f_x$, where $\langle u \rangle$ is the local mean streamwise fluid velocity at the particle center and $f_x=1.7$ is O'Neill's wall-correction factor \cite{stokesEffectInternalFriction1851,oneillSphereContactPlane1968}. The aerodynamic lift force follows Mollinger's formulation \cite{mollingerMeasurementLiftForce1996}, $F_L = 5.3\pi\rho_f\nu^2(d_p^+/2)^{1.87}f_y$, where $d_p^+$ is the dimensionless particle diameter in wall units and $f_y=3.39$ is Maude's wall-correction factor \cite{maudeMovementSphereFront1963}. The mean adhesion force is calculated via the Rabinovich approximation for the interaction potential between a sphere and a rough substrate, $F_{a}=(A_Hd_p/12z_0^2)[1/(1+29d_pR_q/a^2)+1/(1+1.82R_q/z_0)^2]$, where $A_H$ is the Hamaker constant, $z_0$ is the distance of closest approach, and $R_q$ is the root-mean-square surface roughness \cite{rabinovichAdhesionNanoscaleRough2000}. The aerodynamic-force parameters are determined from by physical properties of the fluid and particle, whereas the adhesion-force parameters are determined by the particle-surface interaction and the surface topography, which can be evaluated using atomic force microscopy \cite{peillonAdhesionForcesRadioactive2022}.

Within this framework, applying Reynolds decomposition separates the overall angular velocity $\omega$ into mean ($\langle \omega \rangle$) and fluctuating ($\omega'$) components. The mean component is governed by the deterministic moment balance in equation \ref{eq:moment_balance}, whereas the fluctuating component is treated as a stochastic process. Consistent with K41 theory \cite{kolmogorovLocalStructureTurbulence1941}, this model evolves via a standard Ornstein-Uhlenbeck process (OUP) \cite{riskenFokkerPlanckEquation1989}:

\begin{equation}
    \mathrm{d}\omega'(t) = -\omega'(t)\frac{\mathrm{d}t}{T_L} + \sqrt{\frac{2\sigma^2}{T_L}} \, \mathrm{d}W(t)
    \label{eq:ornstein_uhlenbeck}
\end{equation}
where $T_L$ represents the macroscopic integral timescale of resuspension, $\sigma$ controls the variance and autocorrelation of the process, and $W(t)$ denotes a standard Wiener process \cite{riskenFokkerPlanckEquation1989}. The stochastic increment is expressed as $\mathrm{d}W(t) = \sqrt{\mathrm{d}t}\xi$, where $\xi$ represents uncorrelated white noise sampled independently at each time step from a Gaussian distribution.

At each numerical integration step, the total angular velocity is evaluated as $\omega (t)=\langle \omega \rangle + \omega'(t)$. The numerical contact rule is implemented as follows: assuming the particle can roll in the downstream direction rolling only  \cite{fuParticleResuspensionWallbounded2013,benshlomoIntroducingSurfaceRoughness2024}, negative values of the total angular velocities used for the displacement update are truncated at zero, representing a stochastic state reset. This operation imposes the unidirectional rolling constraint but does not reinitialize the underlying stochastic variables: $\omega'$ and, in the second-order formulation, tangential-acceleration $\alpha'$ are advanced continuously from one integration step to the next until detachment. The detachment criterion is based on the nominal downstream rolling displacement over one viscous wall time $T_0 = \nu / u_\tau^2$. For a particle of diameter $d_p$, this displacement is $\Delta x=(d_p / 2)\omega \Delta t$. A trajectory is classified as detached at the first time for which $\Delta x \geq a$, or equivalently  $\omega_c \geq \omega_c = 2a / (d_p \Delta t)$. Once this condition is satisfied, the trajectory is terminated and removed from the attached-particle ensemble. If it is not satisfied, the trajectory continues from its current dynamical state. This threshold represents operational loss of the initial particle–surface contact over one characteristic asperity spacing; it does not resolve subsequent lift-off, re-entrainment, or transport into the bulk flow. The event is therefore referred to below as detachment rather than complete resuspension. The viscous wall time $T_0$ is used as the threshold-evaluation and nondimensionalization scale. It is not the integral time of the angular-velocity process. The prescribed relaxation time is $T_L=k_0T_0$, with $k_0=100$, so that $T_L$ corresponds to 100 viscous time units \cite{henryParticleResuspensionChallenges2023}.

The variance parameter is defined as $\sigma = \sqrt{2 C_0 T_0 \langle M'^2 \rangle} / I$, where $C_{0}$ is a free calibration parameter acting as a phenomenological surrogate for the aerodynamic memory induced by near-wall coherent structures \cite{ben-shlomoMarkovianAssumptionNearwall2026}. The variance of the fluctuating aerodynamic moment, $\langle M'^2 \rangle$, is expressed as $\langle M'^2 \rangle=(6\pi\rho_f\nu(d_p/2))^2[b^2f_x^2\langle u^2 \rangle + (ab/2)f_xf_y\langle uv \rangle + (a/2)^2f_y^2\langle v^2 \rangle]$, where $\langle u^2 \rangle$, $\langle v^2 \rangle$, and $\langle uv \rangle$ are the local fluid Reynolds stresses evaluated at the particle's position, with $u$ and $v$ denoting the streamwise and wall-normal velocity components, respectively. Substituting $\sigma$ into Equation~\eqref{eq:ornstein_uhlenbeck} yields the explicit stochastic differential equation for the fluctuating angular velocity:

\begin{equation}
    \mathrm{d}\omega'(t) = -\omega'(t)\frac{\mathrm{d}t}{T_L} + \frac{2}{I} \sqrt{\frac{C_0}{k_0}} \left[ 6\pi \rho_f \nu \left(\frac{d_p}{2}\right) \right] \sqrt{ b^2 f_x^2 \langle u^2 \rangle + \frac{ab}{2} f_x f_y \langle uv \rangle + \left(\frac{a}{2}\right)^2 f_y^2 \langle v^2 \rangle } \, \mathrm{d}W(t)
    \label{eq:ornstein_uhlenbeck_explicit}
\end{equation}
equation \ref{eq:ornstein_uhlenbeck_explicit} is integrated numerically at each time step to resolve $\omega'(t)$, completing the first-order resuspension modeling framework.

The second-order formulation retains the deterministic mean angular velocity $\langle \omega \rangle$ obtained from the moment balance in equation \ref{eq:moment_balance}, but enlarges the stochastic state from the angular-velocity fluctuation $\omega'$ alone to the coupled state $(\omega',\alpha')$. This construction is inspired by, but is not a direct implementation of, the Lagrangian velocity–acceleration model developed by Innocenti et al. \cite{innocentiLagrangianStochasticModelling2020} for modelling fluid acceleration in wall-bounded turbulence. Their formulation evolves the joint state of fluid-particle position, velocity, and acceleration, in an inhomogeneous turbulent flow. In the present work, the formulation is reduced to the single local degree of freedom relevant to unidirectional particle rolling. The spatial evolution of the carrier-flow state, the tensorial acceleration dynamics, and the Reynolds averaged Navier Stokes (RANS)-PDF coupling employed by Innocenti et al. \cite{innocentiLagrangianStochasticModelling2020} are therefore not retained.

The rationale for introducing acceleration as an additional state variable follows the time-scale arguments underlying second-order Lagrangian stochastic models \cite{innocentiLagrangianStochasticModelling2020,sawfordReynoldsNumberEffects1991,popeStochasticLagrangianModel2002}. A first-order position-velocity description is appropriate when the acceleration correlation time is much shorter than the observation and velocity-relaxation time scales, so that acceleration can be represented as delta-correlated forcing \cite{sawfordReynoldsNumberEffects1991,popeStochasticLagrangianModel2002}. This separation may become inadequate at low Reynolds numbers and in the near-wall region, where the characteristic time scale of the small-scale dynamics can become comparable to the time scales governing the resolved motion \cite{innocentiLagrangianStochasticModelling2020}. Retaining acceleration as an independent state variable then allows the fluctuating forcing to possess a finite correlation time while preserving a Markovian description in an augmented state space \cite{innocentiLagrangianStochasticModelling2020,sawfordReynoldsNumberEffects1991,popeStochasticLagrangianModel2002}. Thus, the coupled process $(\omega',\alpha')$ is Markovian, although the angular-velocity process $\omega'$, considered alone, is temporally correlated.

To obtain the present one-dimensional reduction, we denote $R=d_p/2$ as the particle radius and define the fluctuating tangential rolling velocity as $U'_r=R\omega'$. Following the general velocity-acceleration structure introduced by Sawford \cite{sawfordReynoldsNumberEffects1991}, subsequently generalized to anisotropic and inhomogeneous turbulent flows by Pope \cite{popeStochasticLagrangianModel2002} and Innocenti et al. \cite{innocentiLagrangianStochasticModelling2020}, the local scalar analogue is written as

\begin{subequations}
\label{eq:ornstein_uhlenbeck_second-order_system_gen}
\begin{empheq}[left=\empheqlbrace]{align}
    \mathrm{d}U_r'(t) &= -U_r'(t)\frac{\mathrm{d}t}{T_L} + \alpha'(t)\mathrm{d}t \label{eq:second-order_velocity_gen} \\[8pt]
    \mathrm{d}\alpha'(t) &= -\alpha'(t)\frac{\mathrm{d}t}{\tau_\eta} + \sqrt{\frac{A_0\langle \epsilon \rangle}{\tau_\eta^2}} \, \mathrm{d}W(t) \label{eq:second-order_acceleration}
\end{empheq}
\end{subequations}

Dividing equation \ref{eq:second-order_velocity_gen} by $R=d_p/2$ gives the coupled equations used here for the particle angular-velocity $(\omega')$ and tangential-acceleration $(\alpha')$ fluctuations:

\begin{subequations}
\label{eq:ornstein_uhlenbeck_second-order_system}
\begin{empheq}[left=\empheqlbrace]{align}
    \mathrm{d}\omega'(t) &= -\omega'(t)\frac{\mathrm{d}t}{T_L} + \frac{2}{d_p}\alpha'(t)\mathrm{d}t \label{eq:second-order_velocity} \\[8pt]
    \mathrm{d}\alpha'(t) &= -\alpha'(t)\frac{\mathrm{d}t}{\tau_\eta} + \sqrt{\frac{A_0\langle \epsilon \rangle}{\tau_\eta^2}} \, \mathrm{d}W(t) \tag{\ref{eq:second-order_acceleration}}
\end{empheq}
\end{subequations}
Here, $\alpha'$ represents an effective local tangential acceleration fluctuation acting on the rolling degree of freedom. It is not intended to reproduce the complete three-dimensional material acceleration of the surrounding fluid. The term $-\omega'/T_L$ represents relaxation of the rolling-velocity fluctuation over the macroscopic time scale $T_L$, whereas equation \ref{eq:second-order_acceleration} represents the more rapidly decorrelating contribution associated with unresolved small-scale forcing \cite{innocentiLagrangianStochasticModelling2020,sawfordReynoldsNumberEffects1991,popeStochasticLagrangianModel2002}. The factor $2/d_p$ converts tangential acceleration into an angular-velocity rate under the assumed rolling kinematics.

The acceleration-relaxation time is identified with the Kolmogorov time scale \cite{innocentiLagrangianStochasticModelling2020,sawfordReynoldsNumberEffects1991}, $\tau_\eta=\sqrt{\nu/\langle\epsilon\rangle}$, where $\langle\epsilon\rangle$ is the mean turbulent-kinetic energy dissipation rate evaluated within the viscous sublayer. The dimensional dissipation rate is prescribed from DNS-based data \cite{tarduWallDissipationRevisited2017a,schlatterAssessmentDirectNumerical2010,zaripovExtremeEventsTurbulent2020} using ⟨$\langle\epsilon\rangle=\langle\epsilon\rangle^+u_\tau^4/\nu$. Combining this relation with the definition of the Kolmogorov time scale gives $\tau_\eta=\sqrt{\nu/\langle\epsilon\rangle}=\sqrt{\nu/(\langle\epsilon\rangle^+u_\tau^4/\nu)}=\nu/u_\tau^2\sqrt{\langle\epsilon\rangle^+}=T_0/\sqrt{\langle\epsilon\rangle^+}$. The separation between the angular-velocity relaxation time and the acceleration-correlation time is therefore $T_L/\tau_\eta=k_0\sqrt{\langle\epsilon\rangle^+}$. For the representative value $\langle\epsilon\rangle^+=0.18$ and $k_0=100$, $\tau_\eta\approx2.36T_0$ and $T_L/\tau_\eta\approx42.4$. Because $k_0$ and $\langle\epsilon\rangle^+$ are held fixed in the calculations, this prescribed time-scale ratio is independent of $Re_\tau$. Even over the tested dissipation range $0.15\leq\langle\epsilon\rangle^+\leq0.24$, $T_L/\tau_\eta$ remains between approximately $38.7$ and $49.0$.

The parameter $A_0$ controls the amplitude of the effective acceleration forcing and is calibrated for the present resuspension model. It is not inherited directly from Innocenti et al. \cite{innocentiLagrangianStochasticModelling2020}, because the modeled variable, dimensional reduction, and calibration target differ from those of their three-dimensional fluid-particle formulation. In particular, Innocenti et al. coupled the velocity–acceleration model to a Reynolds-averaged Navier–Stokes solver to represent an inhomogeneous turbulent channel flow \cite{innocentiLagrangianStochasticModelling2020}, whereas the present model applies the reduced acceleration process locally to the tangential forcing experienced by a wall-adjacent rolling particle.

The reduced formulation assumes a statistically stationary local mean flow and locally prescribed stochastic coefficients. It therefore does not claim to reproduce the complete inhomogeneous velocity-acceleration field or the full consistency properties of the original RANS-PDF formulation \cite{innocentiLagrangianStochasticModelling2020,popeStochasticLagrangianModel2002}. Instead, it is used to test whether introducing a finite acceleration-correlation time changes the pre-detachment rolling dynamics and their higher-order statistics. The time-scale basis of this reduction requires the acceleration dynamics to remain faster than the rolling-velocity relaxation, $\tau_\eta<T_L$ \cite{innocentiLagrangianStochasticModelling2020,moninStatisticalFluidMechanics1979}. Under the present parameterization, $T_L/\tau_\eta\approx42.4$, and the imposed separation is therefore maintained throughout the investigated Reynolds-number range. This ratio is a property of the selected model parameters and should not be interpreted as a Reynolds-number-dependent prediction. Any Reynolds-number dependence observed in the simulated particle statistics arises from changes in the aerodynamic forcing, the particle and threshold quantities expressed in wall units, the surface interaction, and the conditioning or termination of trajectories at detachment, rather than from a change in the prescribed value of $T_L/\tau_\eta$.

The central innovation of the second-order formulation is therefore a change in the temporal structure of the stochastic trajectory rather than a simple increase in the number of state variables. In the first-order model, white noise acts directly on $\omega'$, producing a continuous but nondifferentiable angular-velocity path. In the second-order model, white noise acts on the acceleration state, and $\omega'$ becomes differentiable over lags shorter than $\tau_\eta$ \cite{riskenFokkerPlanckEquation1989,innocentiLagrangianStochasticModelling2020,sawfordReynoldsNumberEffects1991,popeStochasticLagrangianModel2002}. This change of regularity introduces a resolved short-time dynamical layer between the stochastic forcing and the macroscopic detachment event. At the transport level, this layer is expected to appear as a finite-time departure from Fickian MSD scaling \cite{zotero-item-1149,metzlerRandomWalksGuide2000}. The multiscale analyses below therefore examine how acceleration-level memory is accumulated into anomalous surface transport and subsequently compressed into the macroscopic detachment response.

\subsection{Numerical conditions and parameter values}\label{sec:Numerics}

The calculations consider spherical particles of diameter $d_p=13 \mu m$ and density $\rho_p=19,300 kg/m^3$, immersed in a fluid of density $\rho_f=1.29kg/m^3$ and kinematic viscosity $\nu=1.51\times10^{-5} m^2/s$. The substrate is characterized by a root-mean-square roughness $R_q=2.2 nm$, a mean peak-to-peak asperity spacing $a=320nm$, and an adhesion-force lever arm $r_a=32.5nm$. The particle–surface interaction is evaluated using a Hamaker constant $A_H=16\times10^{-20} J$ and a closest separation distance $z_0=0.3 nm$. Unless otherwise specified, these dimensional properties are held fixed across all friction Reynolds numbers investigated.

For each $Re_\tau$, the friction velocity $u_\tau$ determines the viscous time scale $T_0=\nu/u_\tau^2$, the particle diameter in wall units $d_p^+=d_pu_\tau/\nu$, and the wall-normal location of the particle relative to the viscous sublayer. The particle-center elevation is denoted by $y_c^+=y_c u_\tau/\nu$, and the elevation of its uppermost point is $y_{top}^+=y_c^++(d_p^+)/2$. For all cases considered, $y_{top}^+<5$. The particle therefore remains fully immersed within the viscous sublayer throughout the investigated range $60\leq Re_\tau \leq430$, and the same rolling-dominated resuspension mechanism can be applied consistently across the simulations.

The particle-scale friction Reynolds number is $Re_{\tau,p}=d_p u_\tau/\nu=d_p^+$. This quantity is distinguished here from the slip-based particle Reynolds number, $Re_p=(d_p \lvert\langle u\rangle-U_p\rvert)/\nu$, where $U_p$ is the particle translational velocity. When the Stokes approximation is applied, the particle relaxation time and corresponding viscous-scale Stokes number are $\tau_p=(\rho_p d_p^2)/(18\rho_f \nu)$ and $St=\tau_p/T_0 =\rho_p/(18\rho_f )(d_p^+)^2$.

Each condition was simulated using $N=1,000$ independent realizations for the resuspension-fraction calculation, the trajectory, autocorrelation, structure-function, and increment-PDF analyses. Each realization was integrated until particle detachment or until the maximum physical time $t_{max}=1s$, whichever occurred first. The numerical integration step is denoted by $\delta t$, to distinguish it from the physical viscous time scale $T_0$ used in the resuspension criterion. The simulations used $\delta t=T_0$, directly coupling the numerical increment to the viscous time scale. Temporal convergence was previously verified by Fu et al. \cite{fuParticleResuspensionWallbounded2013} through sensitivity analyses using different time steps $(0.1T_0<\delta t<100T_0)$, confirming that $\delta t=T_0$ is the optimal balance between numerical stability and temporal resolution. The initial conditions were $x(0)=0$, $\omega' (0)=0$, and, for the second-order model, $\alpha' (0)=0$.

The parameters common to all conditions are $f_x=1.7$, $f_y=3.39$, $k_0=100$, $C_0=1\times10^{-3}$ for the first-order model, and $A_0=2\times10^{-6}$ for the second-order model. Unless otherwise stated, the second-order calculations use the representative normalized dissipation rate $\langle\epsilon\rangle^+=0.18$.

\subsection{Autocorrelation and structure functions}\label{sec:ACF_Structure_Func}

To quantify the multi-scale statistics and intermittency of the fluctuations, as well as to evaluate the performance of the stochastic models, we utilize temporal autocorrelation functions (ACF), structure functions, and increment kurtosis.
Let $\omega'_i(t) = \omega_i(t) - \langle \omega_i \rangle$ denote the stationary, zero-mean fluctuation of interest at time $t$, where $i$ represents the spatial coordinate index.
The dimensionless temporal autocorrelation function, $R_{ii}(\tau)$, as a function of the lag time $\tau$, is defined by normalizing the two-time autocovariance by the variance of the signal, $\langle \omega'_i(t)^2 \rangle$:
\begin{equation}
    R_{ii}(\tau) = \frac{\langle \omega'_i(t) \omega'_i(t + \tau) \rangle}{\langle \omega'_i(t)^2 \rangle}, \quad \text{where} \quad R_{ii}(0) = 1.
    \label{eq:autocorr_normalized}
\end{equation}
The autocorrelation function characterizes the persistence of the fluctuations over time, with its integral over all lag times defining the macroscopic integral timescale of the process \cite{pressNumericalRecipes3rd2007}.

To further resolve the statistical distribution of the fluctuating velocity increments, we evaluate the temporal structure functions. The general $p$-th order temporal structure function, $S_p(\tau)$, is defined as the ensemble average of the $p$-th power of the increments over a time lag $\tau$:
\begin{equation}
    S_p(\tau) = \langle [\omega'_i(t + \tau) - \omega'_i(t)]^p \rangle.
    \label{eq:structure_function_general}
\end{equation}
In the context of stochastic modeling and second-order closures, the second-order structure function ($p = 2$), denoted here as $D_{ii}(\tau)$, is of primary physical significance as it quantifies the kinetic energy contained within velocity fluctuations of a given time scale:
\begin{equation}
    D_{ii}(\tau) = S_2(\tau) = \langle [\omega'_i(t + \tau) - \omega'_i(t)]^2 \rangle.
    \label{eq:second_order_structure_function}
\end{equation}

$S_2(\tau)$ quantifies the mean-square change in angular velocity over lag $\tau$ and is related to the autocovariance for a stationary process. In classical turbulence theory, $D_{ii}(\tau)$ within the inertial subrange characterizes the rate of turbulent kinetic energy transfer, scaling linearly with the mean dissipation rate in Lagrangian frames ($\sim \langle \epsilon \rangle \tau$). In the limit of vanishingly small lag times ($\tau \to 0$), $D_{ii}(\tau)$ characterizes the local smoothness and dissipation rate of the trajectory, whereas for large lag times exceeding the integral timescale ($\tau \to \infty$), where $R_{ii}(\tau) \to 0$, $D_{ii}(\tau)$ asymptotically plateaus at twice the total variance of the signal ($2\langle\omega'_i(t)^2 \rangle$) \cite{falkovichParticlesFieldsFluid2001,popeTurbulentFlows2000}. The expected small-lag behavior depends on the differentiability of the modeled angular velocity. For the first-order Ornstein--Uhlenbeck process,
$S_2(\tau) = 2\sigma_\omega^2 \left( 1 - e^{-|\tau|/T_L} \right) \simeq \left( \frac{2\sigma_\omega^2}{T_L} \right) |\tau| \quad$ at $|\tau| \ll T_L$, which means that $S_2(\tau) \propto |\tau|$. In the acceleration-augmented formulation, the angular velocity is differentiable over lags shorter than the acceleration-correlation time. Using $\omega(\tau)' \simeq \left( \frac{2}{d_p} \right) \alpha'(t)\tau$ gives $S_2(\tau) \simeq \frac{4}{d_p^2} \langle \alpha'^2 \rangle \tau^2 \quad$ for $|\tau| \ll \tau_\eta$. Thus, the second-order formulation predicts a quadratic, rather than cubic, small-lag angular-velocity structure function.

To characterize the statistical asymmetry and directional bias of the trajectory increments, we evaluate the third-order structure function ($p = 3$):
\begin{equation}
    S_3(\tau) = \langle [\omega'_i(t + \tau) - \omega'_i(t)]^3 \rangle.
    \label{eq:third_order_structure_function}
\end{equation}

While even-order structure functions describe the symmetric magnitude of angular-velocity increments, the third-order structure function, $S_3(\tau)$, retains their sign and therefore measures directional asymmetry. A negative value of $S_3(\tau)$ indicates that negative angular-velocity increments are statistically more intense, more frequent, or both, than positive increments over the corresponding lag time. In the present wall-constrained particle model, such an asymmetry may arise from the deterministic resisting moments, the numerical treatment of particle--surface contact, termination of trajectories at detachment, survivor conditioning, or temporal asymmetry in the applied forcing. The present $S_3(\tau)$ should not be interpreted as a direct measure of the turbulent kinetic-energy cascade. Classical cascade relations, including the four-fifths law, refer to longitudinal spatial velocity increments of the fluid under specific assumptions of homogeneity and isotropy \cite{popeTurbulentFlows2000}. They do not transfer directly to a one-dimensional temporal signal describing the rolling angular velocity of a wall-adjacent particle. Accordingly, $S_3(\tau)$ is used here as a measure of directional asymmetry in the modeled pre-detachment motion. Although a nonzero value may be consistent with temporally asymmetric dynamics, establishing time irreversibility would require a dedicated forward--backward statistical measure and is outside the scope of the present analysis.

Finally, to quantify statistical intermittency and the emergence of intense, heavy-tailed fluctuations, we evaluate the kurtosis (or flatness factor) of the increments, denoted here as $K_{ii}(\tau)$. By normalizing the fourth-order structure function ($p = 4$) by the square of the second-order structure function, the increment kurtosis is defined as:
\begin{equation}
    K_{ii}(\tau) = \frac{S_4(\tau)}{[S_2(\tau)]^2} = \frac{\langle [\omega'_i(t + \tau) - \omega'_i(t)]^4 \rangle}{\langle [\omega'_i(t + \tau) - \omega'_i(t)]^2 \rangle^2}.
    \label{eq:increment_kurtosis}
\end{equation}
The increment kurtosis quantifies departures from Gaussian angular-velocity statistics. For centered Gaussian increments, $K_{ii}(\tau) = 3$, whereas $K_{ii}(\tau) > 3$ indicates a heavier-tailed increment distribution \cite{popeTurbulentFlows2000,frischTurbulenceLegacyKolmogorov1995,xuOriginHighKurtosis1996}. In physical turbulent boundary layers, elevated short-lag kurtosis is commonly associated with intermittent velocity gradients and dissipation \cite{popeTurbulentFlows2000,frischTurbulenceLegacyKolmogorov1995,xuOriginHighKurtosis1996}. Equations \ref{eq:ornstein_uhlenbeck} and \ref{eq:second-order_velocity}-\ref{eq:second-order_acceleration} define the stochastic evolution of the fluctuating angular velocity and acceleration. The ACFs, structure functions, and increment PDFs reported below are evaluated from these fluctuating quantities rather than from the deterministic adhesion moment appearing in equation \ref{eq:moment_balance}. Accordingly, these statistics characterize the temporal organization of the modeled stochastic fluctuations and, where applicable, the conditioning associated with the pre-detachment trajectory records. The second-order formulation is distinguished by the finite correlation time of the acceleration state, which reorganizes the short-time regularity, temporal asymmetry, and increment statistics relative to the first-order formulation.


\section{Results and discussion}\label{sec:results}

\subsection{Macroscopic degeneracy under resuspension-fraction calibration}\label{sec:calibration}

\begin{figure}
	\centering
	\includegraphics[width=0.8\textwidth]{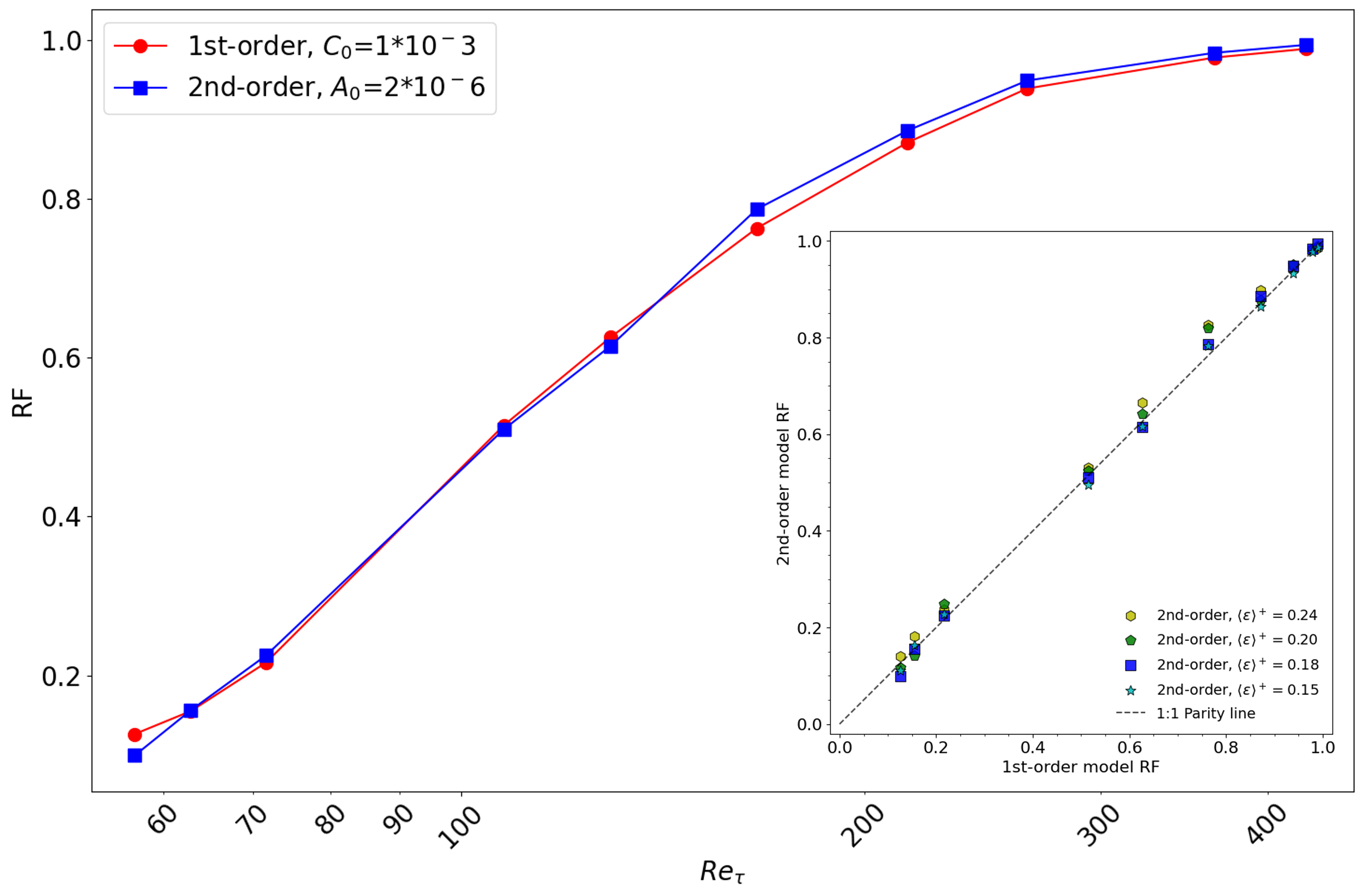}
	\caption{{Resuspended fractions of the first-order model (red circles) and the second-order model (blue squares) as a function of the friction Reynolds number. The inset shows the sensitivity of the resuspended fraction calculation to the normalized kinetic energy dissipation rate value within the viscous sublayer. \label{fig:rf_accel}}}
\end{figure}

Figure~\ref{fig:rf_accel} reveals a near-collapse of the resuspended fractions predicted by the first-order model with $C_0 = 1 \times 10^{-3}$ and the second-order model with $A_0 = 2 \times 10^{-6}$. The value of $A_0$ was selected by calibrating second-order model predictions against the baseline first-order RF until optimal alignment between the two frameworks was obtained. This near-collapse establishes a macroscopic degeneracy: two stochastic formulations with different state spaces and short-time dynamics can be calibrated to the same bulk resuspension response. Figure~\ref{fig:rf_accel} therefore creates the controlled contrast underlying the remainder of the study: nearly identical macroscopic RF curves may emerge from different temporal-memory structures and different anomalous transport histories.

This degeneracy provides a stringent basis for evaluating the information carried by trajectory-level statistics. The first-order formulation has previously reproduced experimental resuspension measurements \cite{benshlomoIntroducingSurfaceRoughness2024}, while the present calibration aligns the second-order formulation with the same macroscopic baseline. The MSD, ACF, structure-function, and PDF analyses then reveal the distinct temporal pathways that remain hidden when the dynamics are reduced to RF.

The two formulations also compress the required flow information differently. The first-order model uses local Reynolds-stress components and $C_0$, whereas the second-order model uses the dissipation rate, acceleration-correlation time, and $A_0$. The second-order formulation therefore replaces several Reynolds-stress inputs with a dissipation-based description while introducing a temporally resolved acceleration state.

The inset of Figure~\ref{fig:rf_accel} examines the sensitivity of the second-order RF prediction to the prescribed normalized dissipation rate, with $A_0$ held fixed at $2 \times 10^{-6}$. The tested range $0.15 \le \langle \epsilon \rangle^+ \le 0.24$ spans representative near-wall values reported in \cite{tarduWallDissipationRevisited2017a,zaripovExtremeEventsTurbulent2020}. Within this range, the macroscopic RF is only weakly affected. Consequently, the subsequent calculations use the representative value $\langle \epsilon \rangle^+ = 0.18$. This is a modeling approximation rather than a universal viscous-sublayer value, because near-wall dissipation can depend on wall-normal position and Reynolds number \cite{tarduWallDissipationRevisited2017a}. Moreover, the weak RF sensitivity does not establish that the ACFs, structure functions, and increment PDFs are equally insensitive to $\langle \epsilon \rangle^+$. The weak sensitivity of RF to $\langle \epsilon \rangle^+$ further emphasizes the central result of Figure~\ref{fig:rf_accel}: macroscopic resuspension is comparatively insensitive to dynamical details that become prominent in multiscale trajectory statistics.

\subsection{Detachment kinetics and anomalous pre-detachment transport}\label{sec:MSD}

\begin{figure}
	\centering
	\includegraphics[width=1.0\textwidth]{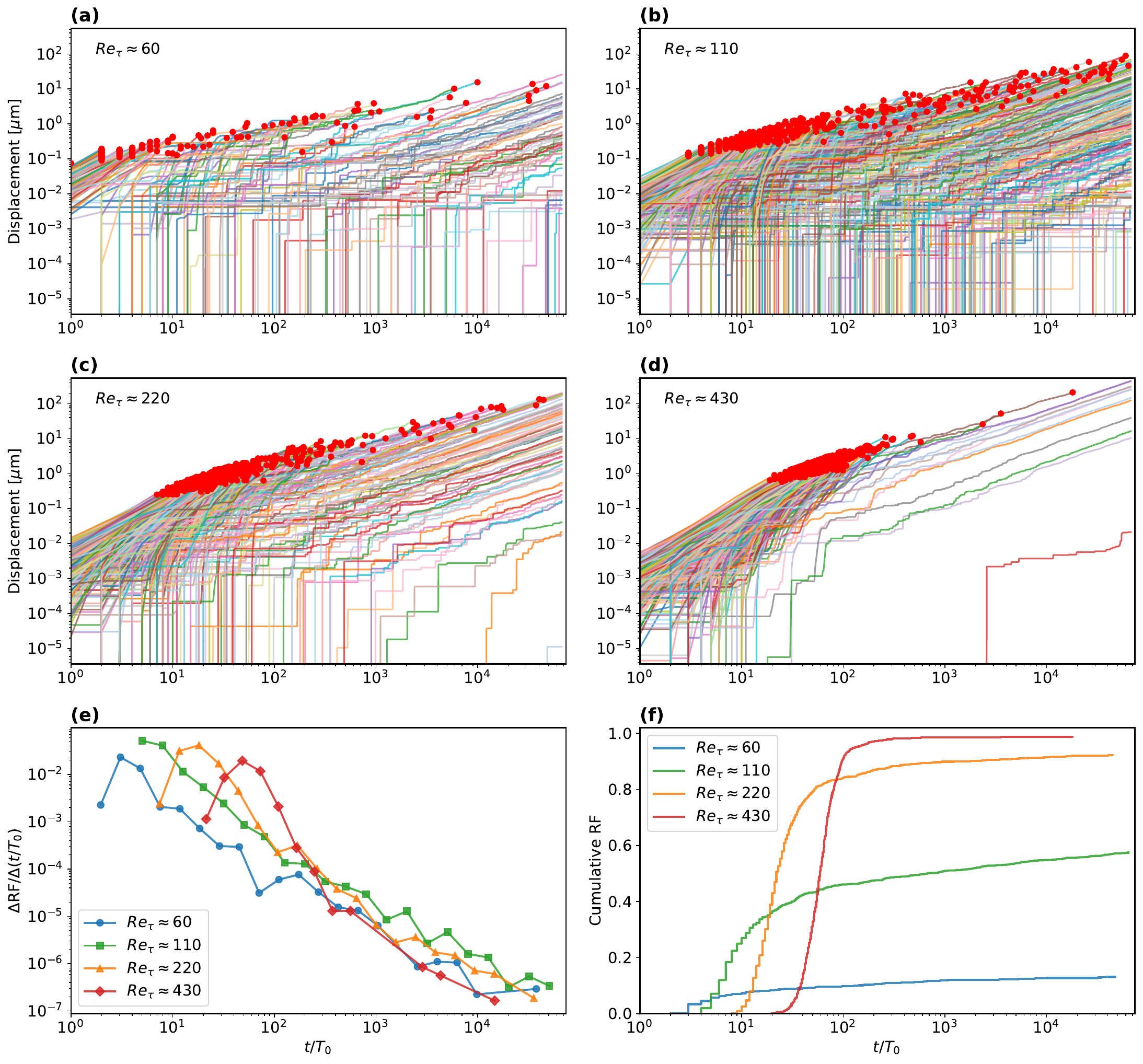}
	\caption{{Particle rolling trajectories and temporal detachment statistics evaluated using the second-order resuspension model across increasing friction Reynolds numbers ($Re_\tau \approx 60$--$430$). (a)-(d) Displacement as a function of normalized time, $t/T_0$, for ensembles of $1{,}000$ particles at $Re_\tau \approx 60$, $110$, $220$, and $430$, respectively. Red circles denote the precise instant and spatial displacement at which individual particles satisfy the operational detachment criterion. (e) Temporal resuspension rate density, $\Delta \text{RF} / \Delta (t / T_0)$, computed using logarithmically spaced time bins. (f) Cumulative resuspended fraction. \label{fig:trajec_2nd}}}
\end{figure}

We analyse the kinematic dynamics of the second-order model by examining the rolling trajectories of particles along the surface as a function of the normalized time, $t/T_0$. Figures~\ref{fig:trajec_2nd}(a)--\ref{fig:trajec_2nd}(d) illustrate the particle-displacement trajectories of $1{,}000$ particles across increasing friction Reynolds numbers ($Re_\tau$). Red markers indicate the time and displacement at which individual particles satisfy the resuspension criterion. Across the investigated Reynolds-number range, most detachment events occur within milliseconds, corresponding approximately to $t/T_0 \sim 10^2$, and therefore well before the end of the prescribed simulation interval \cite{henryParticleResuspensionChallenges2023,fuParticleResuspensionWallbounded2013}. As $Re_\tau$ increases, particles explore progressively larger surface distances before satisfying the detachment criterion, indicating that the route to detachment changes systematically even as the final event remains defined by the same threshold.

\begin{figure}
\centering
\includegraphics[width=1.0\textwidth]{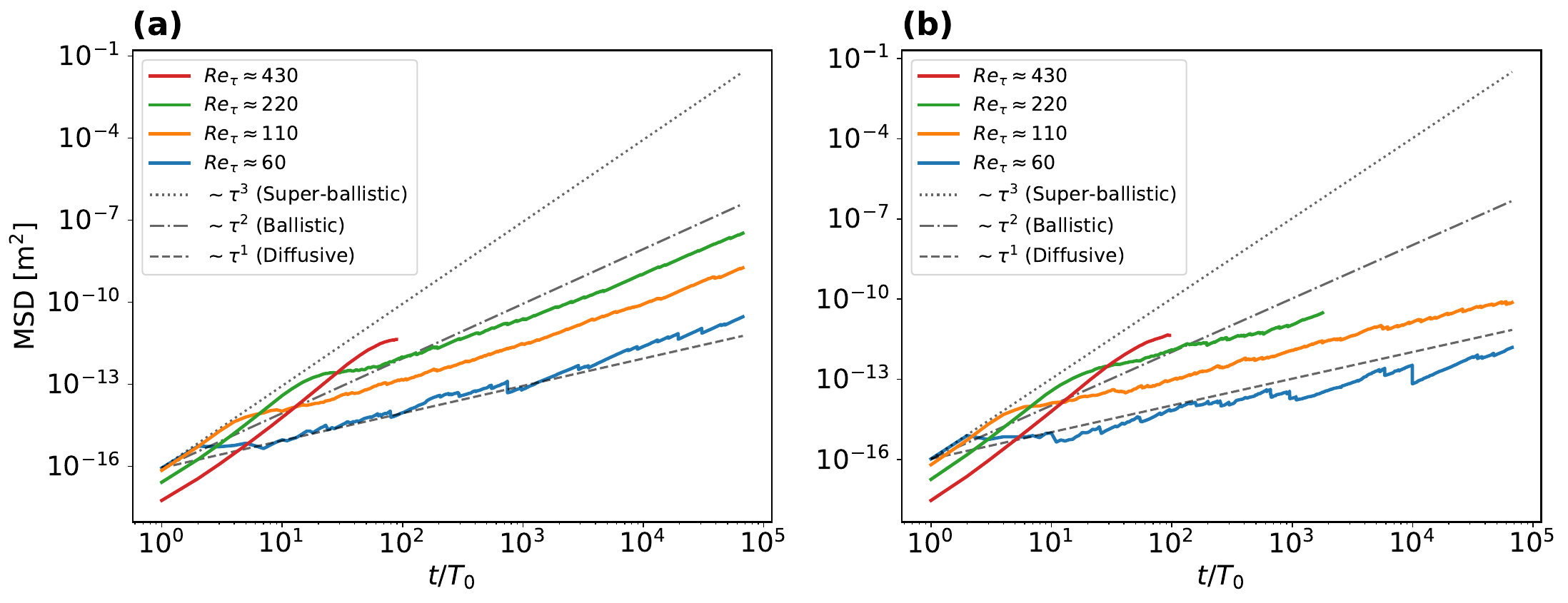}
\caption{{Survivor-conditioned mean-squared displacement, $\mathrm{MSD}_{\mathrm{surv}}(t)$, of particles that remain attached to the surface, plotted as a function of normalized time $t/T_0$ for friction Reynolds numbers $Re_\tau \approx 60$ (blue), $110$ (orange), $220$ (green), and $430$ (red). (a)~First-order resuspension model. (b)~Second-order resuspension model. Reference slopes proportional to $t$, $t^2$, and $t^3$ mark the diffusive, ballistic, and super-ballistic scaling classes used to identify the transient anomalous-transport window \cite{zotero-item-1149,metzlerRandomWalksGuide2000}. Each curve is terminated when the attached population falls below $100$ particles.
\label{fig:MSD}}}
\end{figure}

Figure~\ref{fig:trajec_2nd}(e) presents the temporal detachment-rate density, $\Delta \mathrm{RF}/\Delta(t/T_0)$, while Figure~\ref{fig:trajec_2nd}(f) shows the cumulative detached fraction. Increasing $Re_\tau$ raises the cumulative plateau and translates the detachment-rate distribution toward larger wall-scaled times. The translation combines the Reynolds-number dependence of $T_0$, $d_p^+$, the threshold $\omega_c$, and the balance between aerodynamic and resisting moments. Thus, Figures~\ref{fig:trajec_2nd}(e) and~\ref{fig:trajec_2nd}(f) define the Reynolds-dependent macroscopic kinetics against which the trajectory-level transport in Figure~\ref{fig:MSD} is examined. The simultaneous growth of the detached fraction and translation of its wall-time distribution highlights two distinct aspects of resuspension: the probability of detachment and the dynamical pathway leading to it. The first is summarized by $\mathrm{RF}$, and the second is resolved through the survivor-conditioned MSD and multiscale statistics.

Because trajectories are terminated once the detachment criterion is satisfied, the quantity evaluated here is a survivor-conditioned mean-squared displacement:
\begin{equation}
\mathrm{MSD}_{\mathrm{surv}}(t) = \frac{1}{N_{\mathrm{surv}}(t)} \sum_{i \in L(t)} [x_i(t) - x_i(0)]^2,
\end{equation}
where $L(t)$ denotes the set of particles that remain attached at time $t$, $N_{\mathrm{surv}}(t) = |L(t)|$, $x_i(t)$ is the longitudinal position of particle $i$, and $x_i(0) = 0$ \cite{zotero-item-1149}. Figures~\ref{fig:MSD}(a) and \ref{fig:MSD}(b) portray the $\mathrm{MSD}_{\mathrm{surv}}(t)$ progression across several friction Reynolds numbers ($Re_\tau$) for the first-order and second-order resuspension models, respectively. It is important to note that as the simulation progresses, cumulative particle detachment causes the surviving sample size $N_{\mathrm{surv}}(t)$ to shrink. To maintain statistical robustness, the MSD evaluation for each Reynolds number is terminated once the surviving population drops below a threshold of $N_{\mathrm{surv}}(t) < 100$.

The survivor-conditioned MSD is the natural transport observable for the population that remains dynamically coupled to the surface. It combines two physical processes: memory-driven rolling of the attached particles and progressive selection through detachment. Rapidly moving trajectories leave the ensemble earlier, while the remaining population continues to explore the surface under the combined action of aerodynamic forcing and resisting surface moments.

Figures~\ref{fig:MSD}(a) and \ref{fig:MSD}(b) reveal a distinct transient anomalous-transport window. At short and intermediate times, the MSD grows more rapidly than the diffusive $t^1$ reference and spans regions associated with ballistic and locally super-ballistic behavior \cite{zotero-item-1149,metzlerRandomWalksGuide2000}. At longer times, the growth becomes progressively shallower and approaches a diffusion-like trend. The resulting trajectory is therefore not described by a single transport exponent, but by a multiscale progression from strongly correlated surface exploration toward weaker, decorrelated motion.

Increasing $Re_\tau$ shifts this anomalous-transport window toward larger values of $t/T_0$. Higher turbulence intensity therefore sustains rapid, correlated surface exploration over a broader interval of wall-scaled time before the attached ensemble enters the shallower-growth regime. The position and breadth of this window reflect the joint Reynolds-number dependence of the aerodynamic moments, $T_0$, $d_p^+$, the detachment threshold, surface resistance, and the evolving survivor population. The MSD thus resolves a dynamical layer that is absent from the final resuspended fraction: $\mathrm{RF}$ records how many particles detach, whereas $\mathrm{MSD}_{\mathrm{surv}}$ records the anomalous transport pathway by which the attached population approaches detachment.

Accordingly, Figure \ref{fig:MSD} identifies a Reynolds-number-dependent anomalous-transport pathway in the attached-particle ensemble: correlated rolling produces a finite-time interval of non-Fickian surface exploration, detachment progressively reorganizes the surviving population, and the combined process evolves toward diffusion-like growth.

\subsection{Velocity and acceleration autocorrelation}\label{sec:autocorrelation}

\begin{figure}
	\centering
	\includegraphics[width=1.0\textwidth]{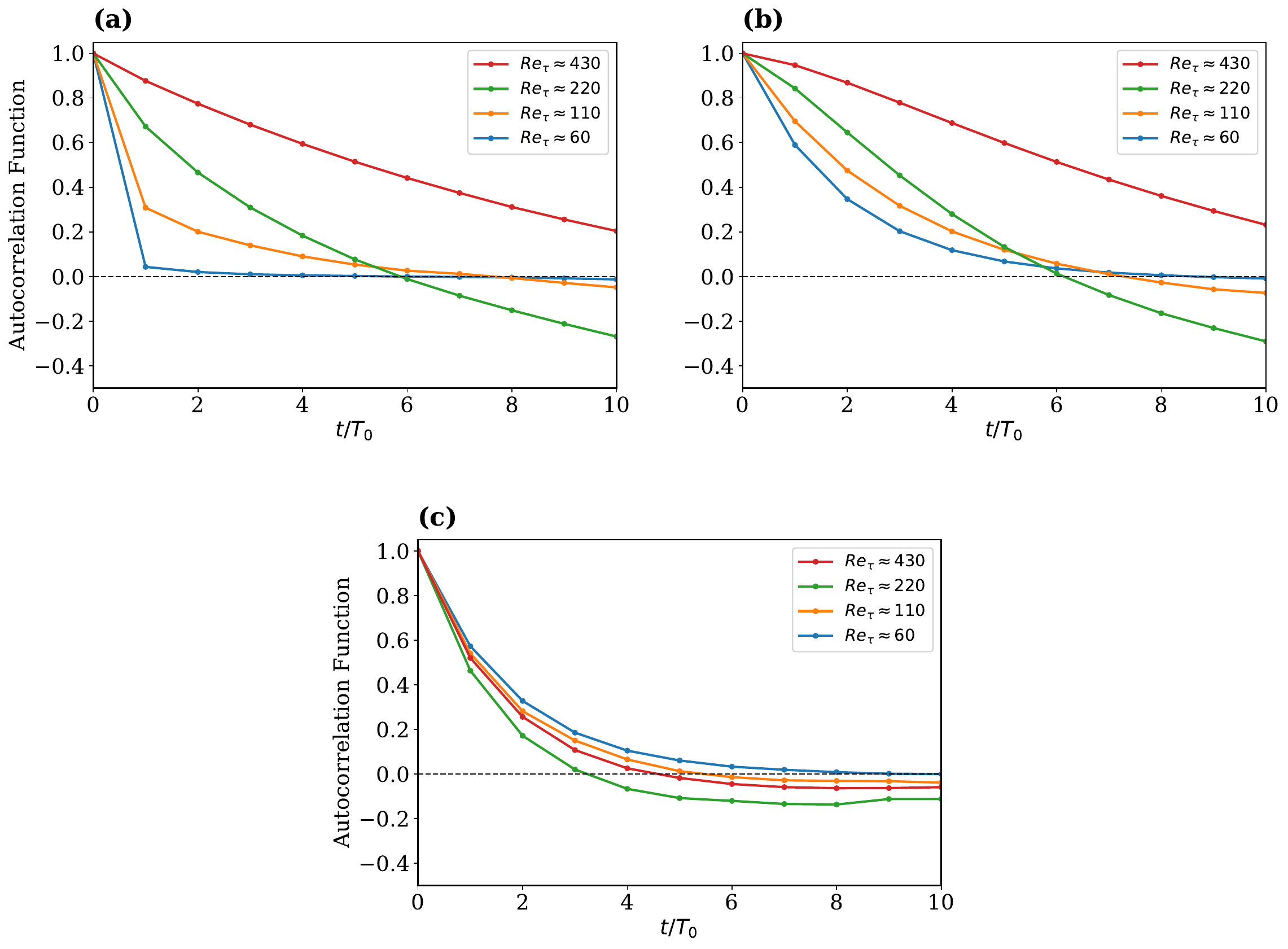}
	\caption{{Ensemble-averaged ACF of the rolling particles as a function of normalized lag time, $t/T_0$, evaluated across friction Reynolds numbers $Re_\tau \approx 60$ (blue), $110$ (orange), $220$ (green), and $430$ (red). (a) Velocity ACF ($R_{\omega'\omega'}$) for the first-order model. (b) Velocity ACF ($R_{\omega'\omega'}$) for the second-order model. (c) Acceleration ACF ($R_{\alpha'\alpha'}$) for the second-order model. \label{fig:ACF}}}
\end{figure}

The autocorrelation functions resolve the temporal-memory architecture of the anomalous-transport window identified in Figure \ref{fig:MSD}. Figures \ref{fig:ACF}(a) and \ref{fig:ACF}(b) show that the first-order angular velocity loses short-lag correlation more rapidly, whereas the acceleration-augmented process retains a smoother and more persistent temporal organization. This distinction follows directly from the level at which stochastic forcing enters the model: white noise acts on $\omega'$ in the first-order formulation and on the finite-correlated acceleration state in the second-order formulation \cite{innocentiLagrangianStochasticModelling2020,sawfordReynoldsNumberEffects1991,popeStochasticLagrangianModel2002}.

Despite their different initial decay rates, the first- and second-order angular-velocity ACFs exhibit similar decorrelation ranges. The corresponding MSD and ACF are mathematically related, but the MSD transition is governed by the accumulated correlation rather than universally by the first zero crossing. For a stationary zero-mean angular-velocity fluctuation and rolling radius $R = d_p/2$, $\langle [x'(t) - x'(0)]^2 \rangle = 2R^2 \int_0^t (t - \tau) C_\omega(\tau) \,\mathrm{d}\tau$ where $C_\omega(\tau) = \langle \omega'(t) \omega'(t+\tau) \rangle$ \cite{popeTurbulentFlows2000,zotero-item-1149}. The integrated correlation therefore controls the evolution of the transport regime, while the first zero crossing provides a practical marker of particle-level decorrelation. Positive accumulated angular-velocity correlation sustains the steep, non-Fickian portion of the MSD, whereas its progressive decay accompanies the crossover toward weaker, diffusion-like growth \cite{popeTurbulentFlows2000,zotero-item-1149}. Figures \ref{fig:MSD} and \ref{fig:ACF} thus connect microscopic temporal memory to the mesoscopic anomalous-transport pathway preceding macroscopic detachment.

Because the ACFs are evaluated directly from the fluctuating rolling signal, their Reynolds-number dependence defines the temporal-memory spectrum of the modeled pre-detachment fluctuations. This spectrum characterizes how the stochastic angular-velocity dynamics reorganize across the investigated Reynolds numbers and provides the dynamical link between the short-time stochastic formulation and the anomalous surface exploration resolved by the MSD.

Secondly, as the friction Reynolds number increases, the velocity ACFs of both models exhibit a slower decay toward zero. This Reynolds-number dependence cannot originate from the prescribed ratio $T_L/\tau_\eta$, because $k_0$ and $\langle \epsilon \rangle^+$ are held fixed and therefore $T_L/\tau_\eta = k_0 \sqrt{\langle \epsilon \rangle^+}$ remains constant. The observed Reynolds-number dependence is therefore a property of the simulated pre-detachment fluctuation ensembles rather than a consequence of a Reynolds-number-dependent change in the prescribed $T_L/\tau_\eta$. It reveals a systematic reorganization of the particle-level temporal correlations across the investigated flow conditions.

Across the cases with $Re_\tau \ge 110$, the second-order acceleration ACF in Figure \ref{fig:ACF}(c) decays more rapidly than the corresponding angular-velocity ACF in Figure \ref{fig:ACF}(b). For example, at $Re_\tau \approx 430$, the angular-velocity ACF remains approximately $0.9$ at $t/T_0 = 2$, whereas the acceleration ACF decreases to approximately $0.25$. This difference is consistent with the role assigned to acceleration as the faster state variable in the augmented stochastic formulation \cite{innocentiLagrangianStochasticModelling2020,sawfordReynoldsNumberEffects1991,popeStochasticLagrangianModel2002}.

At $Re_\tau \approx 60$, the acceleration and angular-velocity ACFs exhibit more similar decay profiles: both decrease to approximately $0.35$ by $t/T_0 = 2$ and approach zero near $t/T_0 \approx 8-10$. This convergence cannot be attributed to a collapse of $T_L/\tau_\eta$, because $k_0$ and $\langle \epsilon \rangle^+$ are fixed and therefore $T_L/\tau_\eta = k_0 \sqrt{\langle \epsilon \rangle^+}$ remains constant. The convergence therefore identifies a distinct low-$Re_\tau$ statistical organization of the modeled angular-velocity and acceleration fluctuations. Its occurrence close to the transitional Reynolds-number range is noteworthy, particularly because intermittent low- and high-drag states have been reported in channel flows at $Re_\tau \approx 70-100$ \cite{agrawalLowHighDragIntermittencies2020}. The present ACF convergence should be reported as a property of the modeled particle trajectories and not as evidence that physical near-wall intermittency or turbulent scale separation disappears.

\subsection{Lagrangian structure functions}\label{sec:structure}

\begin{figure}
	\centering
	\includegraphics[width=1.0\textwidth]{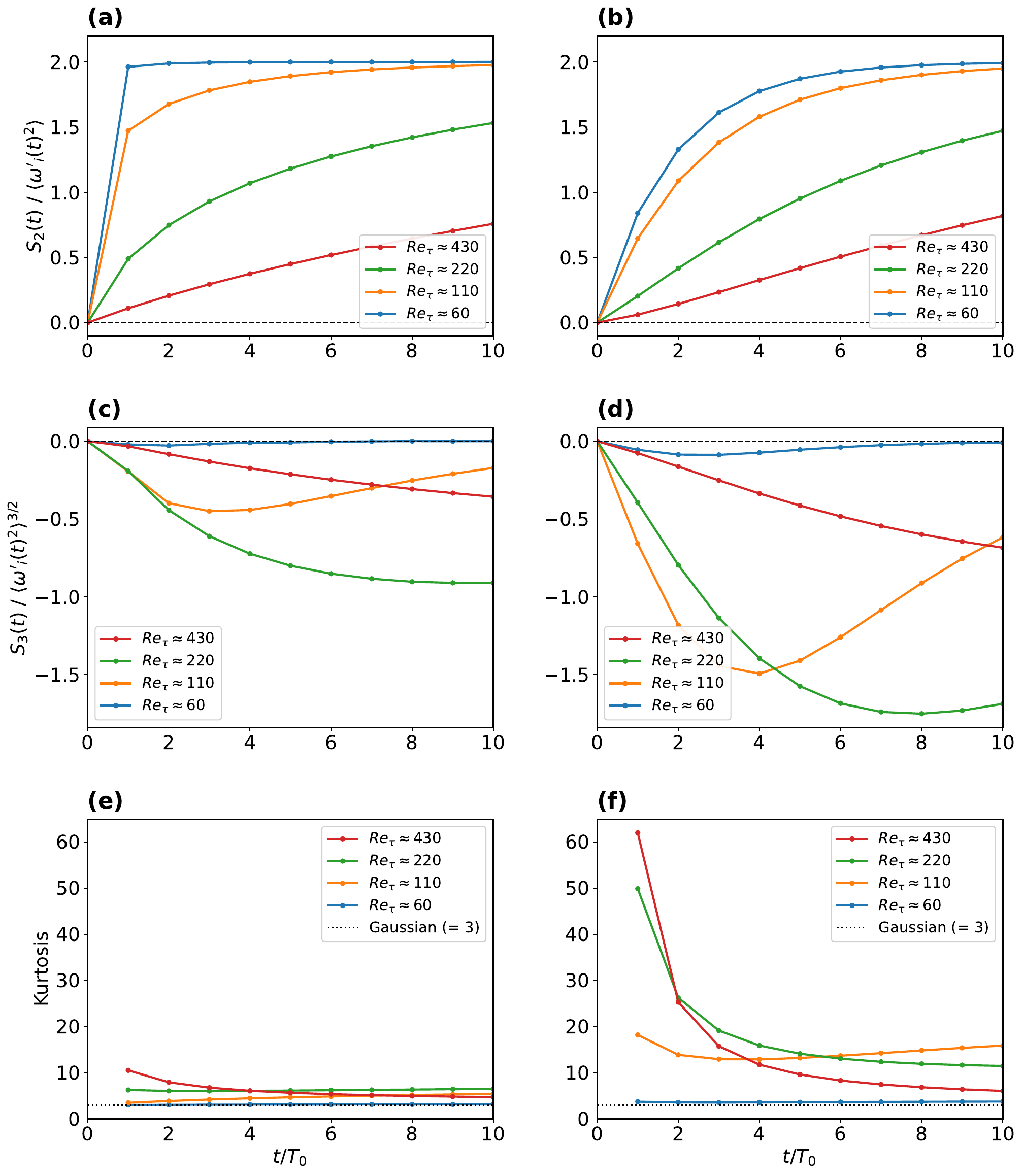}
	\caption{Lagrangian structure functions of the rolling angular velocity evaluated across friction Reynolds numbers $Re_\tau \approx 60$ (blue), $110$ (orange), $220$ (green), and $430$ (red). Second-order structure functions ($S_2$) normalized by the global variance $\langle \omega'_i(t)^2 \rangle$ for the (a) first-order and (b) second-order models. Third-order structure functions ($S_3$) normalized by $\langle \omega'_i(t)^2 \rangle^{3/2}$ for the (c) first-order and (d) second-order models. Kurtosis ($S_4/S_2^2$) for the (e) first-order and (f) second-order models. \label{fig:structure_functions}}
\end{figure}

The most direct signature of dynamical non-equivalence appears in the second-order structure function. To further characterize the intermittency and temporal scaling of the particle's rolling dynamics, the second-, third-, and fourth-order (kurtosis) Lagrangian structure functions of the angular velocity increments were evaluated. Figure \ref{fig:structure_functions} portrays these structure functions for both the first- and second-order resuspension models. For a stationary process with variance $\sigma_\omega^2$, the second-order structure function and ACF are related by ${S_2(\tau)}/{\sigma_\omega^2} = 2 \left[ 1 - R_{\omega\omega}(\tau) \right]$. The normalized structure function therefore begins at zero and approaches two as the angular-velocity correlation vanishes \cite{popeTurbulentFlows2000}. A value of two corresponds to $R_{\omega\omega} = 0$, but it does not by itself define the transition of the survivor-conditioned MSD to a diffusive regime. The primary difference between the two models occurs at short lag.

The different initial slopes follow from the regularity of the two modeled processes. In the first-order formulation, white noise acts directly on $\omega'$, producing a nondifferentiable Ornstein-Uhlenbeck process for which $S_2(\tau) \propto \tau$ at small lag. In the second-order formulation, $\omega'$ is driven by an acceleration process with finite variance and finite correlation time. Consequently, $\omega(\tau)' \simeq \left( \frac{2}{d_p} \right) \alpha' \tau$ and $S_2(\tau) \propto \tau^2$ for $\tau \ll \tau_\eta$. The transition from linear to quadratic short-lag scaling is a structural consequence of introducing finite acceleration correlation. It represents a change in the geometry of the modeled trajectory, from a rough angular-velocity path to a temporally smooth one, and provides a model-order fingerprint that remains hidden in RF. The $S_2$ and MSD results therefore expose complementary levels of the same dynamical novelty: $S_2$ identifies the local regularity class of the angular-velocity trajectory, whereas $\mathrm{MSD}_{\mathrm{surv}}$ reveals how that regularity accumulates into anomalous surface transport before detachment \cite{popeTurbulentFlows2000,zotero-item-1149,metzlerRandomWalksGuide2000}. At larger lags, both structure functions evolve toward their decorrelated limits, carrying the short-time regularity distinction into the broader anomalous-transport crossover observed in Figure \ref{fig:MSD}. At $Re_\tau \approx 430$, rapid detachment narrows the observable short-time window and brings the two curves closer together, illustrating how macroscopic removal kinetics can mask the underlying regularity distinction.

Beyond trajectory regularity, the third-order structure function reveals directional organization in the route to detachment. The negative $S_3$ values in Figures \ref{fig:structure_functions}(c) and \ref{fig:structure_functions}(d) show that negative angular-velocity increments dominate the strongest asymmetric events. Their magnitude increases and their characteristic lag shifts with $Re_\tau$, demonstrating a Reynolds-dependent reorganization of pre-detachment motion. The acceleration-augmented formulation maintains this negative asymmetry over a broader lag interval than the first-order formulation. Finite acceleration memory therefore reorganizes the temporal sequence of angular-velocity fluctuations and maintains the negative asymmetry over a broader lag interval than in the first-order formulation. The resulting $S_3$ provides a trajectory-level signature of dynamical non-equivalence: two formulations producing nearly the same RF, nevertheless generate markedly different temporal asymmetry in the fluctuations preceding detachment.

The increment kurtosis $(S_4/S_2^2)$ in Figures \ref{fig:structure_functions}(e) and \ref{fig:structure_functions}(f) quantifies the breadth of the pre-detachment trajectory ensemble. For Gaussian increments $K = 3$, whereas $K > 3$ identifies heavy-tailed statistics enriched in intense events \cite{popeTurbulentFlows2000,frischTurbulenceLegacyKolmogorov1995,xuOriginHighKurtosis1996}. At the higher Reynolds numbers, the second-order formulation produces substantially larger short-lag kurtosis than the first-order formulation. Finite acceleration memory is therefore associated with a substantially broader spectrum of short-lag angular-velocity increments, providing a second higher-order signature of the dynamical distinction between the two formulations. As lag increases, the kurtosis decreases and the trajectory ensemble progressively loses its small-scale concentration of extreme increments. This decay provides the distributional counterpart of the ACF and MSD crossovers: temporal memory, anomalous surface transport, directional asymmetry, and heavy-tailed increments evolve over overlapping multiscale intervals. Together, these observables reveal the structured pathway to detachment that remains invisible in RF.

At $Re_\tau \approx 60$, the trajectory statistics enter a distinct low-Reynolds-number state in which the velocity and acceleration ACFs converge and the increment kurtosis remains near three. Because $T_L/\tau_\eta$ is fixed, this Gaussianization represents a change in the statistical character of the modeled fluctuation ensemble rather than a change in the prescribed scale separation. It therefore identifies a particle-level statistical transition near the onset of sustained turbulence, alongside the low- and high-drag intermittencies reported in channel flows at $Re_\tau \approx 70-100$ \cite{agrawalLowHighDragIntermittencies2020}.

\begin{figure}
	\centering
	\includegraphics[width=0.8\textwidth]{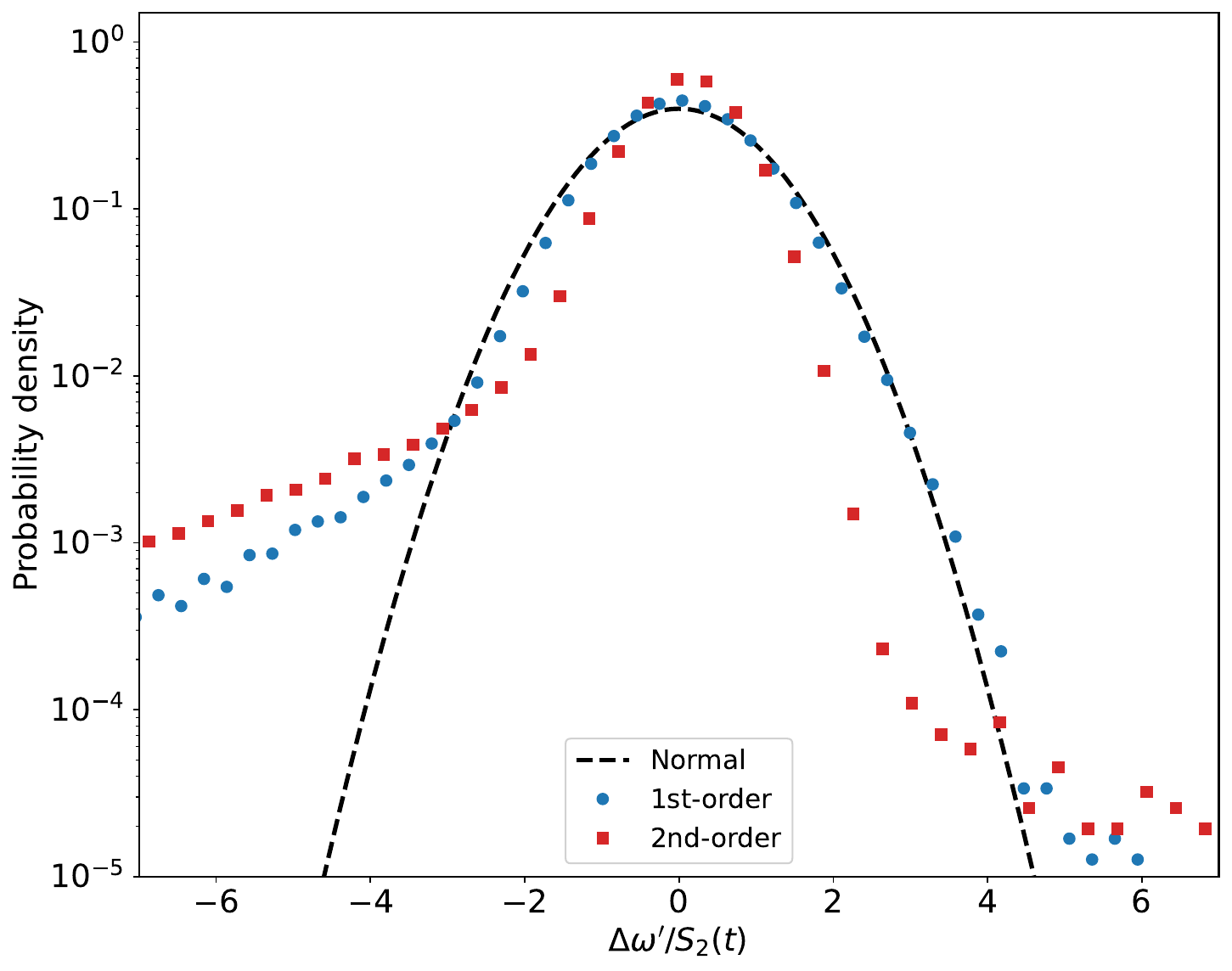}
	\caption{PDF of the fluctuating angular velocity increments evaluated at $t/T_0 = 1.0$ for the highest turbulence intensity ($Re_\tau \approx 430$) investigated. The distributions generated by the first-order model (blue circles) and second-order model (red squares) are plotted on a logarithmic scale and compared against a Standard Normal Gaussian baseline (dashed black line). \label{fig:PDF}}
\end{figure}

To verify the non-Gaussian statistics captured by the third- and fourth-order structure functions, we evaluated the PDF of the standardized fluctuating angular velocity increments, $\Delta \omega' / \sqrt{S_2(t)}$, at a characteristic short time lag ($t/T_0 = 1.0$). Figure \ref{fig:PDF} portrays these distributions at the highest investigated turbulence intensity ($Re_\tau \approx 430$), plotted on a logarithmic scale against a standard Normal (Gaussian) baseline. On this coordinate system, a purely Gaussian stochastic process manifests as a parabola. As illustrated in Figure \ref{fig:PDF}, both modelling frameworks deviate from the Gaussian reference and exhibit a pronounced asymmetry toward negative angular-velocity increments. The broader negative tail of the second-order distribution provides a direct distributional counterpart to the negative $S3$ and elevated kurtosis shown in Figure \ref{fig:structure_functions}. In particular, the acceleration-augmented formulation generates a larger population of intense negative increments than the first-order model. Together, Figures \ref{fig:structure_functions} and \ref{fig:PDF} demonstrate that introducing finite acceleration correlation changes not only the regularity and autocorrelation of the trajectory, but also its directional and rare-event statistics. These differences remain largely invisible in the macroscopic RF results, reinforcing the dynamical non-equivalence of the two formulations.


\section{Conclusions}~\label{sec:conclusions}

A second-order Lagrangian stochastic model was developed to resolve the pre-detachment dynamics of wall-adjacent particles by augmenting angular velocity with a finite-correlated tangential acceleration. Comparison with an established first-order model over $60 \le Re_\tau \le 430$ reveals a striking separation between macroscopic and trajectory-level behavior. After calibration, the two formulations produce nearly identical resuspended fractions, establishing a macroscopic degeneracy between distinct stochastic dynamics.

Multiscale trajectory statistics break this degeneracy. The finite acceleration-correlation time changes the short-time regularity of the angular velocity from $S_2(\tau) \propto \tau$ to $S_2(\tau) \propto \tau^2$ for $\tau \ll \tau_\eta$, converting a rough Ornstein-Uhlenbeck path into a differentiable short-time trajectory. This model-order fingerprint propagates across scales: the second-order model retains angular-velocity memory over a broader short-lag interval, sustains stronger directional asymmetry, and generates heavier increment tails.

The survivor-conditioned MSD provides the transport-level expression of this temporal restructuring. Figure \ref{fig:MSD} reveals a finite-time anomalous-transport window in which the attached-particle ensemble explores the surface more rapidly than the diffusive reference and locally spans ballistic and super-ballistic scaling classes \cite{zotero-item-1149,metzlerRandomWalksGuide2000}. The subsequent crossover toward diffusion-like growth records the progressive loss of angular-velocity memory together with the reorganization of the attached population through detachment. Increasing $Re_\tau$ shifts this anomalous window toward larger wall-scaled times, demonstrating that Reynolds number changes not only the probability of detachment but also the temporal pathway by which particles approach it. At $Re_\tau \approx 60$, the convergence of velocity and acceleration decorrelation and the approach toward Gaussian increments identify a distinct particle-level low-Reynolds-number state \cite{patelObservationsSkinFriction1969,hamiltonRegenerationMechanismsNearwall1995,agrawalLowHighDragIntermittencies2020}.

The principal result is that a correct macroscopic resuspended fraction does not uniquely determine the dynamics that produce it. Finite acceleration correlation introduces a trajectory-resolved layer of information that is compressed out of RF: short-time regularity, temporal memory, anomalous surface transport, directional organization, and rare-event statistics. The anomalous-transport window provides the missing mesoscopic link between stochastic forcing and bulk detachment kinetics; it records how particles reach the threshold, whereas RF records only how many do so. These findings establish multiscale trajectory observables as essential discriminants for stochastic resuspension models and provide a framework for validating both macroscopic detachment and the dynamical pathways leading to it.

\section*{Acknowledgement}

The authors of this work are grateful for the generous support of Pazy Foundation, grant ID 133-2020 and the John R. Goldsmith Memorial Prize Fund.

\bibliography{better_bib}

@article{agrawalLowHighDragIntermittencies2020,
  title = {Low- and {{High-Drag Intermittencies}} in {{Turbulent Channel Flows}}},
  author = {Agrawal, Rishav and Ng, Henry C.-H. and Davis, Ethan A. and Park, Jae Sung and Graham, Michael D. and Dennis, David J.C. and Poole, Robert J.},
  year = 2020,
  month = oct,
  journal = {Entropy},
  volume = {22},
  number = {10},
  pages = {1126},
  doi = {10.3390/e22101126},
  urldate = {2025-01-12},
  copyright = {https://creativecommons.org/licenses/by/4.0/},
  langid = {english}
}

@article{ben-shlomoMarkovianAssumptionNearwall2026,
  title = {On the {{Markovian}} Assumption in Near-Wall Turbulence: The Case of Particle Resuspension},
  shorttitle = {On the {{Markovian}} Assumption in Near-Wall Turbulence},
  author = {{Ben-Shlomo}, David and Berkovich, Ronen and Fattal, Eyal},
  year = 2026,
  month = apr,
  journal = {Journal of Fluid Mechanics},
  volume = {1033},
  pages = {R1},
  doi = {10.1017/jfm.2026.11435},
  urldate = {2026-04-15},
  langid = {english}
}

@article{benshlomoIntroducingSurfaceRoughness2024,
  title = {Introducing Surface Roughness in Adhesion for Stochastic and {{Rock}}'n'{{Roll}} Models to Describe Particle Resuspension in Turbulent Flows},
  author = {Ben Shlomo, David and Almog, Roy and Klausner, Ziv and Fattal, Eyal and Berkovich, Ronen},
  year = 2024,
  month = may,
  journal = {Surfaces and Interfaces},
  volume = {48},
  pages = {104321},
  doi = {10.1016/j.surfin.2024.104321},
  urldate = {2024-06-22}
}

@article{falkovichParticlesFieldsFluid2001,
  title = {Particles and Fields in Fluid Turbulence},
  author = {Falkovich, G. and Gaw{\c e}dzki, K. and Vergassola, M.},
  year = 2001,
  month = nov,
  journal = {Reviews of Modern Physics},
  volume = {73},
  number = {4},
  pages = {913--975},
  doi = {10.1103/RevModPhys.73.913},
  urldate = {2026-07-20}
}

@article{fattalHeterogenousCanopyLagrangianStochastic2023,
  title = {Heterogenous {{Canopy}} in a {{Lagrangian-Stochastic Dispersion Model}} for {{Particulate Matter}} from {{Multiple Sources}} over the {{Haifa Bay Area}}},
  author = {Fattal, Eyal and {David-Saroussi}, Hadas and Buchman, Omri and Tas, Eran and Klausner, Ziv},
  year = 2023,
  month = jan,
  journal = {Atmosphere},
  volume = {14},
  number = {1},
  pages = {144},
  doi = {10.3390/atmos14010144},
  urldate = {2023-07-29},
  copyright = {http://creativecommons.org/licenses/by/3.0/},
  langid = {english}
}

@book{frischTurbulenceLegacyKolmogorov1995,
  title = {Turbulence: {{The Legacy}} of {{A}}. {{N}}. {{Kolmogorov}}},
  shorttitle = {Turbulence},
  author = {Frisch, Uriel},
  year = 1995,
  month = nov,
  langid = {english}
}

@article{fuParticleResuspensionWallbounded2013,
  title = {Particle Resuspension in a Wall-Bounded Turbulent Flow},
  author = {Fu, S. C. and Chao, C. Y.H. and So, R. M.C. and Leung, W. T.},
  year = 2013,
  journal = {Journal of Fluids Engineering, Transactions of the ASME},
  volume = {135},
  number = {4},
  doi = {10.1115/1.4023660}
}

@article{guingoNewModelSimulation2008,
  title = {A New Model for the Simulation of Particle Resuspension by Turbulent Flows Based on a Stochastic Description of Wall Roughness and Adhesion Forces},
  author = {Guingo, Mathieu and Minier, Jean Pierre},
  year = 2008,
  journal = {Journal of Aerosol Science},
  volume = {39},
  number = {11},
  doi = {10.1016/j.jaerosci.2008.06.007}
}

@article{hamiltonRegenerationMechanismsNearwall1995,
  title = {Regeneration Mechanisms of Near-Wall Turbulence Structures},
  author = {Hamilton, James M. and Kim, John and Waleffe, Fabian},
  year = 1995,
  month = mar,
  journal = {Journal of Fluid Mechanics},
  volume = {287},
  pages = {317--348},
  doi = {10.1017/S0022112095000978},
  urldate = {2026-07-22},
  langid = {english}
}

@article{henryNumericalStudyAdhesion2012,
  title = {Numerical Study on the Adhesion and Reentrainment of Nondeformable Particles on Surfaces: {{The}} Role of Surface Roughness and Electrostatic Forces},
  author = {Henry, Christophe and Minier, Jean Pierre and Lef{\`e}vre, Gr{\'e}gory},
  year = 2012,
  month = jan,
  journal = {Langmuir},
  volume = {28},
  number = {1},
  pages = {438--452},
  doi = {10.1021/la203659q}
}

@article{henryParticleResuspensionChallenges2023,
  title = {Particle Resuspension: {{Challenges}} and Perspectives for Future Models},
  shorttitle = {Particle Resuspension},
  author = {Henry, Christophe and Minier, Jean-Pierre and Brambilla, Sara},
  year = 2023,
  month = mar,
  journal = {Physics Reports},
  series = {Particle Resuspension: Challenges and Perspectives for Future Models},
  volume = {1007},
  pages = {1--98},
  doi = {10.1016/j.physrep.2022.12.005},
  urldate = {2023-05-14},
  langid = {english}
}

@article{henryStochasticApproachSimulation2014,
  title = {A Stochastic Approach for the Simulation of Particle Resuspension from Rough Substrates: {{Model}} and Numerical Implementation},
  shorttitle = {A Stochastic Approach for the Simulation of Particle Resuspension from Rough Substrates},
  author = {Henry, Christophe and Minier, Jean-Pierre},
  year = 2014,
  month = nov,
  journal = {Journal of Aerosol Science},
  volume = {77},
  pages = {168--192},
  doi = {10.1016/j.jaerosci.2014.08.005},
  urldate = {2023-01-06},
  langid = {english}
}

@article{huModelingResuspensionSmall2023,
  title = {Modeling the Resuspension of Small Inertial Particles in Turbulent Flow over a Fractal-like Multiscale Rough Surface},
  author = {Hu, Ruifeng and Johnson, Perry L. and Meneveau, Charles},
  year = 2023,
  month = feb,
  journal = {Physical Review Fluids},
  volume = {8},
  number = {2},
  pages = {024304},
  doi = {10.1103/PhysRevFluids.8.024304},
  urldate = {2023-07-11}
}

@article{ibrahimMicroparticleDetachmentSurfaces2004,
  title = {Microparticle Detachment from Surfaces Exposed to Turbulent Air Row: Microparticle Motion after Detachment},
  author = {Ibrahim, A H and Brach, R M and Dunn, P F},
  year = 2004,
  journal = {Journal of Aerosol Science},
  volume = {35},
  pages = {1189--1204},
  doi = {10.1016/j.jaerosci.2004.05.003}
}

@article{innocentiLagrangianStochasticModelling2020,
  title = {Lagrangian Stochastic Modelling of Acceleration in Turbulent Wall-Bounded Flows},
  author = {Innocenti, Alessio and Mordant, Nicolas and Stelzenmuller, Nick and Chibbaro, Sergio},
  year = 2020,
  month = jun,
  journal = {Journal of Fluid Mechanics},
  volume = {892},
  pages = {A38},
  doi = {10.1017/jfm.2020.203},
  urldate = {2022-12-23},
  langid = {english}
}

@article{jiangCharacterizingEffectSubstrate2008,
  title = {Characterizing the Effect of Substrate Surface Roughness on Particle-Wall Interaction with the Airflow Method},
  author = {Jiang, Yanbin and Matsusaka, Shuji and Masuda, Hiroaki and Qian, Yu},
  year = 2008,
  month = sep,
  journal = {Powder Technology},
  volume = {186},
  number = {3},
  pages = {199--205},
  doi = {10.1016/j.powtec.2007.11.041}
}

@article{kolmogorovLocalStructureTurbulence1941,
  title = {Local Structure of Turbulence in an Incompressible Viscous Fluid at Very Large {{Reynolds}} Numbers},
  author = {Kolmogorov, Andrej Nikolaevich},
  year = 1941,
  journal = {Dokl. Akad. Nauk SSSR},
  volume = {30},
  number = {4},
  pages = {299--301},
  urldate = {2023-06-07}
}

@article{maudeMovementSphereFront1963,
  title = {The Movement of a Sphere in Front of a Plane at Low {{Reynolds}} Number},
  author = {Maude, A. D.},
  year = 1963,
  month = dec,
  journal = {British Journal of Applied Physics},
  volume = {14},
  number = {12},
  pages = {894},
  doi = {10.1088/0508-3443/14/12/316},
  urldate = {2023-07-20},
  langid = {english}
}

@article{metzlerRandomWalksGuide2000,
  title = {The Random Walk's Guide to Anomalous Diffusion: A Fractional Dynamics Approach},
  shorttitle = {The Random Walk's Guide to Anomalous Diffusion},
  author = {Metzler, Ralf and Klafter, Joseph},
  year = 2000,
  month = dec,
  journal = {Physics Reports},
  volume = {339},
  number = {1},
  pages = {1--77},
  doi = {10.1016/S0370-1573(00)00070-3},
  urldate = {2026-07-23}
}

@article{mollingerMeasurementLiftForce1996,
  title = {Measurement of the Lift Force on a Particle Fixed to the Wall in the Viscous Sublayer of a Fully Developed Turbulent Boundary Layer},
  author = {Mollinger, A. M. and Nieuwstadt, F. T. M.},
  year = 1996,
  month = jun,
  journal = {Journal of Fluid Mechanics},
  volume = {316},
  pages = {285--306},
  doi = {10.1017/S0022112096000547},
  urldate = {2021-03-13}
}

@book{moninStatisticalFluidMechanics1979,
  title = {Statistical {{Fluid Mechanics}}},
  author = {Monin, A. S. and Yaglom, A. M.},
  year = 1979,
  volume = {1}
}

@incollection{obukhovDescriptionTurbulenceTerms1959,
  title = {Description of {{Turbulence}} in {{Terms}} of {{Lagrangian Variables}}},
  booktitle = {Advances in {{Geophysics}}},
  author = {Obukhov, A. M.},
  editor = {Landsberg, H. E. and Van Mieghem, J.},
  year = 1959,
  month = jan,
  volume = {6},
  pages = {113--116},
  doi = {10.1016/S0065-2687(08)60098-9},
  urldate = {2023-07-20},
  langid = {english}
}

@article{oneillSphereContactPlane1968,
  title = {A Sphere in Contact with a Plane Wall in a Slow Linear Shear Flow},
  author = {O'Neill, M. E.},
  year = 1968,
  journal = {Chemical Engineering Science},
  volume = {23},
  number = {11},
  doi = {10.1016/0009-2509(68)89039-6}
}

@article{patelObservationsSkinFriction1969,
  title = {Some Observations on Skin Friction and Velocity Profiles in Fully Developed Pipe and Channel Flows},
  author = {Patel, V. C. and Head, M. R.},
  year = 1969,
  month = aug,
  journal = {Journal of Fluid Mechanics},
  volume = {38},
  number = {1},
  pages = {181--201},
  doi = {10.1017/S0022112069000115},
  urldate = {2026-07-22},
  langid = {english}
}

@article{peillonAdhesionForcesRadioactive2022,
  title = {Adhesion Forces of Radioactive Particles Measured by the {{Aerodynamic Method}}--{{Validation}} with {{Atomic Force Microscopy}} and Comparison with Adhesion Models},
  author = {Peillon, Samuel and G{\'e}lain, Thomas and Payet, Micka{\"e}l and Gensdarmes, Fran{\c c}ois and Grisolia, Christian and Pluchery, Olivier},
  year = 2022,
  month = sep,
  journal = {Journal of Aerosol Science},
  volume = {165},
  pages = {106037},
  doi = {10.1016/j.jaerosci.2022.106037},
  urldate = {2023-05-14},
  langid = {english}
}

@article{popeStochasticLagrangianModel2002,
  title = {A Stochastic {{Lagrangian}} Model for Acceleration in Turbulent Flows},
  author = {Pope, Stephen B.},
  year = 2002,
  month = jul,
  journal = {Physics of Fluids},
  volume = {14},
  number = {7},
  pages = {2360--2375},
  doi = {10.1063/1.1483876},
  urldate = {2026-02-12}
}

@book{popeTurbulentFlows2000,
  title = {Turbulent {{Flows}}},
  author = {Pope, Stephen B.},
  year = 2000,
  month = aug,
  doi = {10.1017/CBO9780511840531},
  urldate = {2022-02-08}
}

@book{pressNumericalRecipes3rd2007,
  title = {Numerical Recipes 3rd Edition: {{The}} Art of Scientific Computing},
  author = {Press, William H and Teukolsky, Saul A and Vetterling, William T and Flannery, Brian P},
  year = 2007
}

@article{rabinovichAdhesionNanoscaleRough2000,
  title = {Adhesion between Nanoscale Rough Surfaces. {{I}}. {{Role}} of Asperity Geometry},
  author = {Rabinovich, Yakov I. and Adler, Joshua J. and Ata, Ali and Singh, Rajiv K. and Moudgil, Brij M.},
  year = 2000,
  month = dec,
  journal = {Journal of Colloid and Interface Science},
  volume = {232},
  number = {1},
  pages = {10--16},
  doi = {10.1006/jcis.2000.7167}
}

@book{riskenFokkerPlanckEquation1989,
  title = {The {{Fokker-Planck Equation}}},
  author = {Risken, Hannes},
  editor = {Haken, Hermann},
  year = 1989,
  series = {Springer {{Series}} in {{Synergetics}}},
  volume = {18},
  address = {Berlin, Heidelberg},
  doi = {10.1007/978-3-642-61544-3},
  urldate = {2022-12-24}
}

@article{sawfordReynoldsNumberEffects1991,
  title = {Reynolds Number Effects in {{Lagrangian}} Stochastic Models of Turbulent Dispersion},
  author = {Sawford, B. L.},
  year = 1991,
  journal = {Physics of Fluids A},
  volume = {3},
  number = {6},
  pages = {1577--1586},
  doi = {10.1063/1.857937}
}

@article{schlatterAssessmentDirectNumerical2010,
  title = {Assessment of Direct Numerical Simulation Data of Turbulent Boundary Layers},
  author = {Schlatter, Philipp and {\"O}rl{\"u}, Ramis},
  year = 2010,
  month = sep,
  journal = {Journal of Fluid Mechanics},
  volume = {659},
  pages = {116--126},
  doi = {10.1017/S0022112010003113},
  urldate = {2025-09-09},
  langid = {english}
}

@article{shengBufferLayerStructures2009,
  title = {Buffer Layer Structures Associated with Extreme Wall Stress Events in a Smooth Wall Turbulent Boundary Layer},
  author = {Sheng, J. and Malkiel, E. and Katz, J.},
  year = 2009,
  month = aug,
  journal = {Journal of Fluid Mechanics},
  volume = {633},
  pages = {17--60},
  doi = {10.1017/S0022112009006934},
  urldate = {2024-07-09},
  langid = {english}
}

@article{shnappTurbulenceobstacleInteractionsLagrangian2020,
  title = {Turbulence-Obstacle Interactions in the {{Lagrangian}} Framework: {{Applications}} for Stochastic Modeling in Canopy Flows},
  shorttitle = {Turbulence-Obstacle Interactions in the {{Lagrangian}} Framework},
  author = {Shnapp, Ron and {Bohbot-Raviv}, Yardena and Liberzon, Alex and Fattal, Eyal},
  year = 2020,
  month = sep,
  journal = {Physical Review Fluids},
  volume = {5},
  number = {9},
  pages = {094601},
  doi = {10.1103/PhysRevFluids.5.094601},
  urldate = {2023-07-20}
}

@article{soltaniDirectNumericalSimulation1995,
  title = {Direct Numerical Simulation of Particle Entrainment in Turbulent Channel Flow},
  author = {Soltani, Mehdi and Ahmadi, Goodarz},
  year = 1995,
  month = mar,
  journal = {Physics of Fluids},
  volume = {7},
  number = {3},
  pages = {647--657},
  doi = {10.1063/1.868587},
  urldate = {2025-04-06}
}

@article{soltaniParticleAdhesionRemoval1994,
  title = {On Particle Adhesion and Removal Mechanisms in Turbulent Flows},
  author = {Soltani, Mehdi and Ahmadi, Goodarz},
  year = 1994,
  month = jan,
  journal = {Journal of Adhesion Science and Technology},
  volume = {8},
  number = {7},
  pages = {763--785},
  doi = {10.1163/156856194X00799},
  urldate = {2025-08-18}
}

@article{stokesEffectInternalFriction1851,
  title = {On the Effect of the Internal Friction of Fluids on the Motion of Pendulums},
  author = {Stokes, George Gabriel},
  year = 1851,
  journal = {Transactions of the Cambridge Philosophical Society},
  volume = {9},
  pages = {8}
}

@article{tarduWallDissipationRevisited2017a,
  title = {Near Wall Dissipation Revisited},
  author = {Tardu, Sedat},
  year = 2017,
  month = oct,
  journal = {International Journal of Heat and Fluid Flow},
  series = {Symposium on {{Experiments}} and {{Simulations}} in {{Fluid Dynamics Research}}},
  volume = {67},
  pages = {104--115},
  doi = {10.1016/j.ijheatfluidflow.2017.03.006},
  urldate = {2026-06-01}
}

@article{thomsonCriteriaSelectionStochastic1987,
  title = {Criteria for the Selection of Stochastic Models of Particle Trajectories in Turbulent Flows},
  author = {Thomson, D. J.},
  year = 1987,
  month = jul,
  journal = {Journal of Fluid Mechanics},
  volume = {180},
  pages = {529--556},
  doi = {10.1017/S0022112087001940},
  urldate = {2023-07-20},
  langid = {english}
}

@article{vanhoutSpatiallyTemporallyResolved2013,
  title = {Spatially and Temporally Resolved Measurements of Bead Resuspension and Saltation in a Turbulent Water Channel Flow},
  author = {{van Hout}, Ren{\'e}},
  year = 2013,
  month = jan,
  journal = {Journal of Fluid Mechanics},
  volume = {715},
  pages = {389--423},
  doi = {10.1017/jfm.2012.525},
  urldate = {2022-12-24},
  langid = {english}
}

@article{wilsonReviewLagrangianStochastic1996,
  title = {Review of {{Lagrangian}} Stochastic Models for Trajectories in the Turbulent Atmosphere},
  author = {Wilson, John D. and Sawford, Brian L.},
  year = 1996,
  month = feb,
  journal = {Boundary-Layer Meteorology},
  volume = {78},
  number = {1},
  pages = {191--210},
  doi = {10.1007/BF00122492},
  urldate = {2023-07-20},
  langid = {english}
}

@article{xuOriginHighKurtosis1996,
  title = {Origin of High Kurtosis Levels in the Viscous Sublayer. {{Direct}} Numerical Simulation and Experiment},
  author = {Xu, C. and Zhang, Z. and {den Toonder}, J. M. J. and Nieuwstadt, F. T. M.},
  year = 1996,
  month = jul,
  journal = {Physics of Fluids},
  volume = {8},
  number = {7},
  pages = {1938--1944},
  doi = {10.1063/1.868973},
  urldate = {2025-09-16}
}

@article{zaripovExtremeEventsTurbulent2020,
  title = {Extreme Events of Turbulent Kinetic Energy Production and Dissipation in Turbulent Channel Flow: Particle Image Velocimetry Measurements},
  shorttitle = {Extreme Events of Turbulent Kinetic Energy Production and Dissipation in Turbulent Channel Flow},
  author = {Zaripov, Dinar and Li, Renfu and Saushin, Ilya},
  year = 2020,
  month = jan,
  journal = {Journal of Turbulence},
  volume = {21},
  number = {1},
  pages = {39--51},
  doi = {10.1080/14685248.2020.1727914},
  urldate = {2026-02-05}
}

@article{ziskindParticleResuspensionSurfaces2006,
  title = {Particle Resuspension from Surfaces: {{Revisited}} and Re-Evaluated},
  author = {Ziskind, Gennady},
  year = 2006,
  month = jan,
  journal = {Reviews in Chemical Engineering},
  volume = {22},
  pages = {1--123},
  doi = {10.1515/REVCE.2006.22.1-2.1}
}

@article{zotero-item-1149,
  title = {Diffusion by Continuous Movements},
  author = {Taylor, Geoffrey I.},
  year = 1921,
  journal = {Proceedings of the london mathematical society},
  volume = {2},
  number = {1},
  pages = {196--212}
}

\end{document}